\documentclass{iau} 
\usepackage{graphicx}

\newcommand{\apj}{ApJ}

\newcommand{\aap}{A\&A}

\newcommand{\mnras}{MNRAS}
\newcommand{\nat}{Nature}
\newcommand{\aapr}{A\&ARv}

\title[Explaining the winds of AGB stars] 
{Explaining the winds of AGB stars: Recent progress}

\author[H{\"o}fner \& Freytag]   
{Susanne H{\"o}fner 
   \and Bernd Freytag
}

\affiliation{
Theoretical Astrophysics, Department of Physics \& Astronomy, 
Uppsala University, \\ 
Box 516, SE-75120 Uppsala, Sweden \\
email: {\tt susanne.hoefner@physics.uu.se}}

\pubyear{2022}
\volume{366}  
\jname{The Origin of Outflows in Evolved Stars}
\editors{L. Decin, A. Zijlstra \& C. Gielen, eds.}
\begin{document}

\maketitle

\begin{abstract}
The winds observed around asymptotic giant branch (AGB) stars are generally attributed to radiation pressure on dust, which is formed in the extended dynamical atmospheres of these pulsating, strongly convective stars. Current radiation-hydrodynamical models can explain many of the observed features, and they are on the brink of delivering a predictive theory of mass loss. 
This review  summarizes recent results and ongoing work on winds of AGB stars, discussing critical ingredients of the driving mechanism, and first results of global 3D RHD star-and-wind-in-a-box simulations. With such models it becomes possible to follow the flow of matter, in full 3D geometry, all the way from the turbulent, pulsating interior of an AGB star, through its atmosphere and dust formation zone into the region where the wind is accelerated by radiation pressure on dust. Advanced instruments, which can resolve the stellar atmospheres, where the winds originate, provide essential data for testing the models.
\end{abstract}

\firstsection 
\section{Introduction}\label{s_intro}

While going through the Asymptotic Giant Branch phase, low- and intermediate mass stars develop winds 
with typical mass loss rates of $10^{-7} - 10^{-5}M_{\odot}/$yr and wind velocities of about $5 - 30$ km/s, which affect their observable properties and their further evolution. AGB stars tend to show a pronounced variability of their luminosities and spectra with periods of about $100 - 1000$ days, attributed to large-amplitude pulsations. 
The pulsations, together with large-scale convective flows, trigger strong radiative shock waves in the stellar atmospheres, which intermittently lift gas to distances where temperatures are low enough to permit dust formation. Radiation pressure on dust grains is assumed to be the driving force behind the massive winds of AGB stars, which turn them into white dwarfs and enrich the interstellar medium with newly-produced chemical elements. 

Considerable efforts have been made to understand the physics of AGB stars, and to develop quantitative models of their dynamical interiors, atmospheres and winds. A recent review by 
\cite{hoefolof18} 
discusses both theoretical and observational aspects of AGB star winds in some detail. Here, we give a brief summary of recent developments, focusing on properties of wind-driving dust grains, stellar pulsation and convection, and the 3D morphology of atmospheres and winds. 

\section{Quantitative dynamic models: Ingredients and applications}\label{s_methods}

Ideally, quantitative models of the dynamical atmospheres and winds should be based on first physical principles, describing the complex interplay of dynamical, radiative and micro-physical processes in AGB stars. By following the flow of matter in full 3D geometry, all the way from the convective, pulsating interior of an AGB star, through its atmosphere and dust formation zone into the region where the wind is accelerated by radiation pressure on dust, a predictive theory of mass loss can be developed. Taking snapshots of the resulting structures of the dynamical atmospheres and winds, synthetic observable properties may be computed and compared to observations, in order to test such models.  
While the most advanced 3D “star-and-wind-in-a-box” models (discussed in Sect.\,\ref{s_3Dwind}) are approaching this ideal, it has to be noted that such numerical simulations are computationally very demanding. 

Current dynamic wind models for AGB stars usually assume spherically symmetric flows, which reduces computation times to a degree that makes it possible to construct large grids, covering the wide range of stellar parameters needed in stellar evolution studies. These models come in two basic categories: 
\begin{itemize}
\item
  steady wind models, assuming time-independent radial outflows, and 
\item 
  time-dependent atmosphere and wind models, accounting for effects of pulsation. 
\end{itemize}
Historically, steady wind models have played a critical role in establishing radiation pressure on dust as the most probable driving mechanism, and in studying the composition of the dust grains by comparing synthetic spectral energy distributions to observed data. They are also widely used to derive wind properties (mass loss rates, radial velocity profiles) from observations. 
Their computational efficiency makes them common tools for stellar population studies, and for predicting dust yields (see, e.g., the talk by Nanni, this conference). However, it has to be kept in mind that the mass loss rate is an input parameter in these  models (set via the starting conditions at the foot point of the steady outflow), not a result. 

\begin{figure}
\centering
\includegraphics[width=13.5 cm]{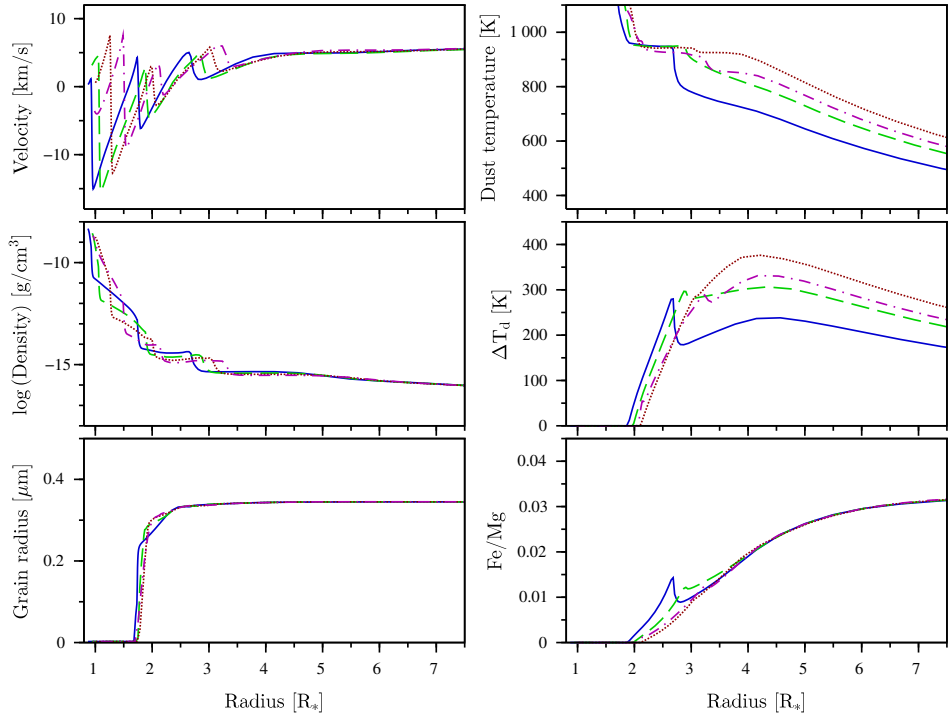}
     \caption{Time-dependent radial structure of a DARWIN model (Dynamical Atmosphere and Radiation-driven Wind model, based on Implicit Numerics; model An315u3 in 
     \cite[H{\"o}fner et al. 2021]{hoefetal21arx}). 
     {\em Left, top to bottom}: Flow velocity, gas density, dust grain radius. 
     {\em Right, top to bottom}: Temperature of the olivine-type Mg-Fe silicate grains, difference in grain temperature with and without taking Fe-enrichment into account, and the Fe/Mg ratio in the dust grains (set by self-regulation via the grain temperature; note the plateaus in dust temperatures, top right panel).  
     The plots show the inner parts of the model, zoomed in on the dust formation and wind acceleration region (snapshots of 4 phases during a pulsation cycle). The first snapshot (solid lines) corresponds to a near-minimum phase, the dashed lines represent the ascending part of the bolometric light curve, the dotted lines shows a phase close to the luminosity maximum, and the dash-dotted lines represent the descending part of the bolometric light curve. It should be noted that only the innermost, dust-free parts of the model structures show periodic variations that repeat every pulsation cycle, while grain growth and wind acceleration are governed by other time scales. 
     The top right panel, showing the highest dust temperatures at maximum light, demonstrates the general effects of radiative heating. The bottom right panel shows the Fe-enrichment, and how this affects the grain temperature is displayed in the middle right panel.
     }
      \label{f_struct}
\end{figure}

In time-dependent models, on the other hand, the physical conditions in the region, where dust forms and initiates an outflow, are a result of pulsation-induced shock waves. These models predict mass loss rates, wind velocities and dust properties for given stellar parameters, elemental abundances and pulsation properties (periods, amplitudes). 
The simulations describe the varying radial profiles of densities, temperatures, velocities, and dust properties (see Fig.\,\ref{f_struct} for an example). These can be used to compute a wide range of synthetic observables, e.g. spectra at various resolutions, photometric fluxes and light curves, interferometric data. Much of our current understanding of dynamical atmospheres and dust-driven winds of AGB stars is derived from such models, as discussed below. 

\section{Properties of wind-driving dust grains}\label{s_dust}

Chemically speaking, AGB stars can be divided into two groups, defined by the value of the atmospheric C/O ratio being smaller (M-type) or larger (C-type) than 1. Due to the high binding energy of the CO molecule, the less abundant of the two elements is almost completely bound up in CO, while the more abundant element is available for forming other molecules and dust. 

In C-type AGB stars (with C/O $>1$ in the atmosphere, due to the 3.\,dredge-up) amorphous carbon is the main wind-driving dust species. The mass loss rates, wind velocities, spectral energy distributions, and photometric variations resulting from time-dependent atmosphere and wind models are in good agreement with observations, see, e.g.,  
\cite{nowoetal11, nowoetal13, eriketal14}. 
However, due to the strong absorption by carbon dust, the stellar photospheres and inner atmospheres may be severely obscured, and detailed comparisons with spatially resolved observations can be difficult, e.g.,  
\cite{palaetal09,stewetal16,sacuetal11,wittetal17}. 

The winds of M-type AGB stars (with C/O $<1$ in the atmosphere) have long been assumed to be driven by radiation pressure on silicate dust, which consists of abundant elements and produces characteristic mid-IR features around 10 and 18 microns, observed in many such stars. Detailed models suggest a scenario where the radiative pressure, that triggers the outflows, is caused by photon scattering on Fe-free silicate grains, 
\cite{hoef08}. 
Such particles are highly transparent at visual and NIR wavelengths, resulting in significantly less radiative heating and smaller condensation distances than for Fe-bearing silicate grains, see 
\cite{dorsetal95,jaegetal03,woit06,zeidetal11}. 
To create sufficient radiative pressure by scattering, the grains need to be of a size comparable to the wavelengths where the stellar flux peaks, meaning that grain radii should fall in the range of 0.1 -- 1$\,\mu$m. Recent observational studies have found dust grains with radii of about 0.1 -- 0.5$\,\mu$m, at distances below 2 -- 3 stellar radii, e.g., 
\cite{norretal12, ohnaetal16, ohnaetal17}, 
as predicted by the numerical simulations.  

Models of winds driven by photon scattering on Fe-free silicate grains
result in visual and near-IR spectra, light curves, variations of photometric colors with pulsation phase, and wind properties that are in good agreement with observations, see  
\cite{bladetal13, bladetal15, bladetal19, hoefetal16}. 
In order to produce the characteristic mid-IR silicate features, however, a gradual enrichment of the silicate dust with Fe in the inner wind region has to be taken into account, to keep grain temperatures from dropping too quickly with distance from the star. 
Figure\,\ref{f_struct} shows new DARWIN models by 
\cite{hoefetal21arx}, 
describing the growth of olivine-type silicate grains with a variable Fe/Mg ratio, set by self-regulation via the grain temperature. The models demonstrate that the  enrichment of the silicate grains with Fe is a secondary process, taking place in the stellar wind on the surface of Fe-free grains that have initiated the outflow. The self-regulating feedback between grain composition and radiative heating leads to low values of Fe/Mg, typically a few percent.

Dust formation in the atmospheres of AGB stars is a non-equilibrium process, since the time-scales of grain growth are comparable to dynamical processes (pulsation, wind acceleration) and variations in radiative flux. Due to falling densities in the outflows, grain growth rates drop strongly with distance from the star, leading to condensation of dust-forming elements that typically is far from complete. In dynamical wind models grain growth is usually treated with a kinetic description, defining the rates at which material is condensing onto the grains out of the surrounding gas. 
A long-standing problem in this context is a realistic description of nucleation, that is the formation of the very first condensation nuclei from the gas. The nature of these first condensates and their formation rates have been a subject of intense debate, see, e.g., \cite{gobretal16,gailetal16} and references therein. In M-type AGB stars, Al$_2$O$_3$ (corundum) is one of the most promising candidates, and 
\cite{gobretal21}
report recent progress in a quantum-chemical description of this species. Condensation of silicates onto pre-existing Al$_2$O$_3$ seed particles may speed up grain growth to sizes necessary for driving a wind by photon scattering, as demonstrated by 
\cite{hoefetal16}. 

\begin{figure}
\centering

\hspace{0.9cm}\includegraphics[width=11.0cm]{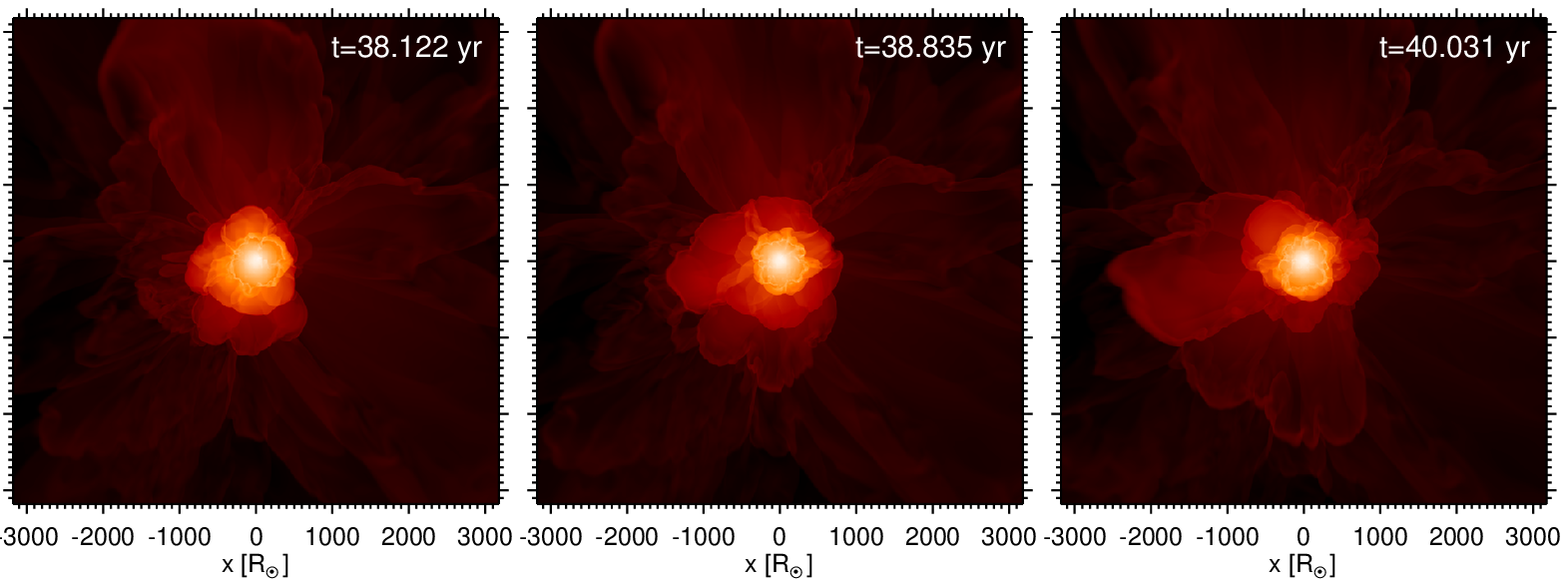}\includegraphics[width=1.36cm]{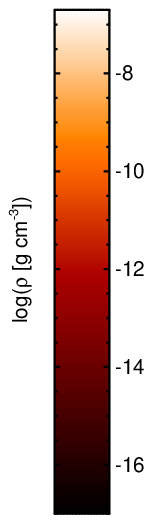}

\hspace{0.9cm}\includegraphics[width=11.0cm]{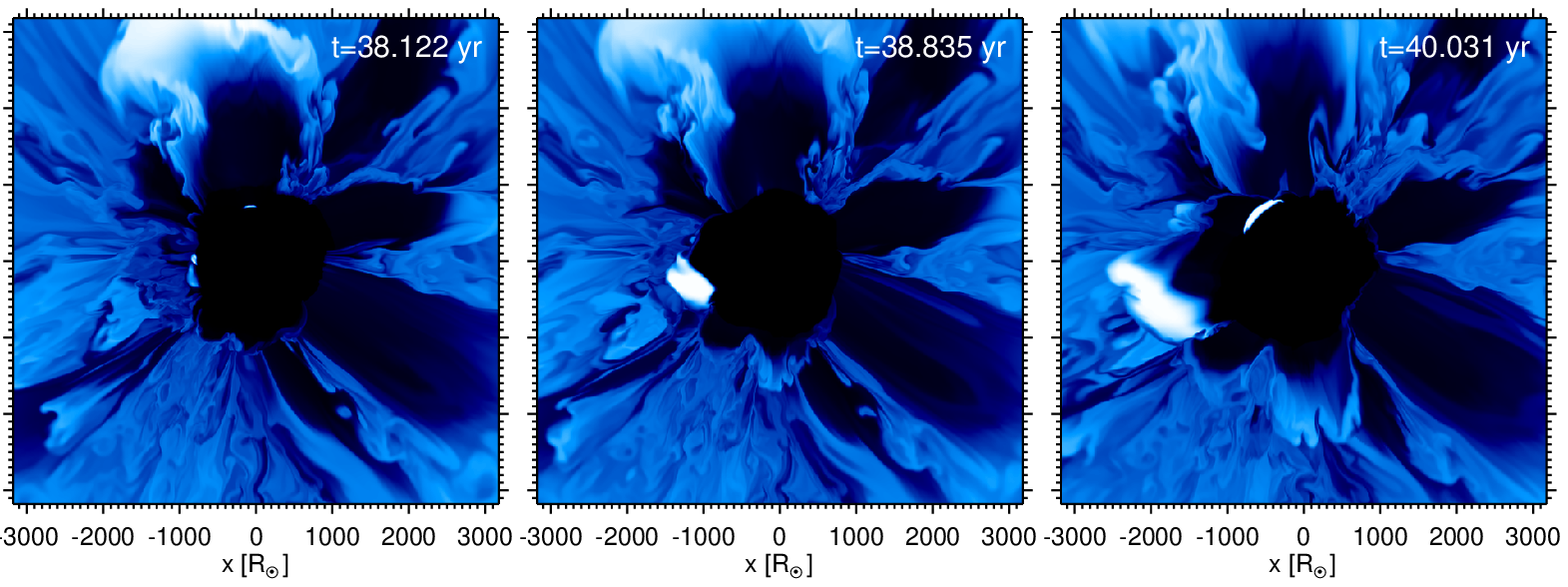}\includegraphics[width=1.36cm]{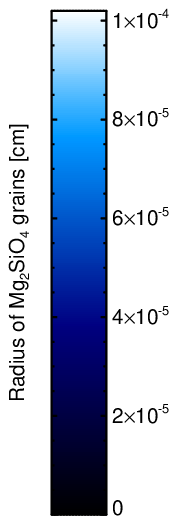}

\hspace{0.9cm}\includegraphics[width=11.0cm]{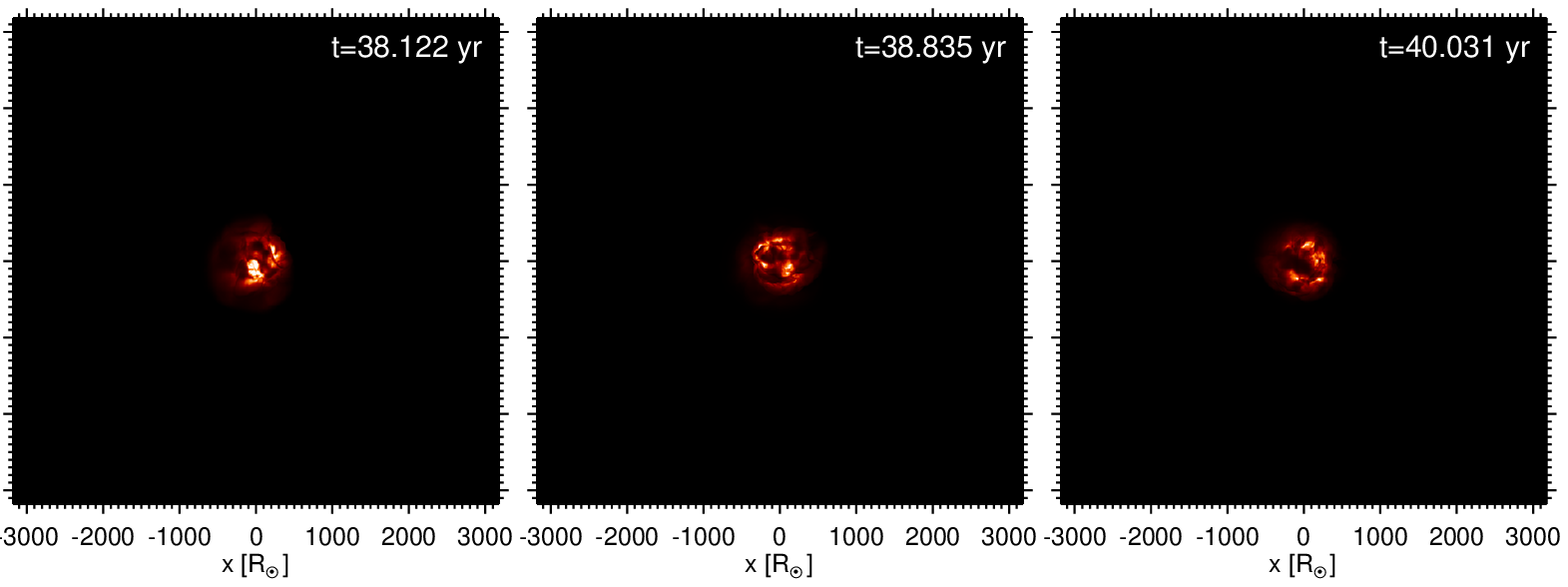}\includegraphics[width=1.36cm]{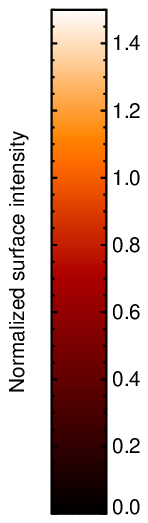}

\caption{Time-dependent structure of a 3D radiation-hydrodynamical model of an AGB star, including interior dynamics (convection, pulsation), dust formation and wind acceleration, computed with the CO5BOLD code (Freytag \& H{\"o}fner, in prep.). 
Time sequences (left to right) of gas density and silicate grain radius for slices through the center of the star (rows 1--2), and the variation of relative surface intensity (bottom row, indicating the size of the star).
The stellar parameters of the model (st28gm06n052) are 1.0\,M$_{\odot}$, 7000\,L$_{\odot}$, and an effective temperature of 2700\,K, and the resulting mass loss rate is about $5 \cdot 10^{-6}\,M_{\odot}$/yr. The sequence of images shows the formation of two new distinct dust clouds (bright areas in the grain size plots, row 2, lower left and upper left quadrants), in the dense wakes of atmospheric shock waves (row 1). At the top edge of the images, a dust cloud that was formed earlier is driven outwards by radiation pressure. 
}\label{f_3Dmod}
\end{figure}

\section{Stellar pulsation and convection}\label{s_puls}

In the current time-dependent atmosphere and wind models discussed above, the effects of radial pulsation are introduced through variable inner boundary conditions, just below the stellar photosphere. In most cases, the periodic expansions and contractions of a star are simulated by prescribing a sinusoidal motion of the innermost layer, accompanied by a variation in luminosity (so-called piston models). The pulsation period may be set according to empirical period--luminosity relations and the amplitude can, in principle, be constrained by comparing the resulting wind properties, synthetic spectra and light curves to observations.  
This simple treatment of pulsation effects has the advantage of few free parameters, but it also has its limitations. Introducing a non-sinusoidal shape of the luminosity variation and a phase shift compared to the motion of the photospheric layers, as suggested by observed light curves and CO line profile variations, may lead to significant differences in mass loss rates and wind velocities, see  
\cite{liljetal17}.  

Ultimatively, a predictive mass loss theory requires realistic models of stellar convection and pulsation, which affect the stellar atmosphere and wind. Pulsations of AGB stars are notoriously difficult to model, but some progress has been made in recent years, regarding lower amplitude overtone pulsations using a linear non-adiabatic approach, e.g.,  
\cite{wood15, trabetal17}. 
Furthermore, 
\cite{trabetal21}
presented new results on non-linear radial pulsation, which resolve earlier problems with predicted fundamental-mode periods, compared to observations of Mira variables. 

A general drawback of 1D pulsation models is that they have to rely on a simplified description of stellar convection. The classical picture of mixing-length style energy transport is probably inadequate 
for the extremely non-linear, non-adiabatic, large-scale interior dynamics of AGB stars. Both theory and observations indicate the existence of giant convection cells, in other words turbulent gas flows on scales comparable to the stellar radius,  
\cite{schw75,freyhoef08,freyetal17,palaetal18}. 
They blur the distinction between convection and pulsation, making the applicability of 1D pulsation models questionable.

A way to resolve these problems is global 3D radiation-hydrodynamical (RHD) modelling. The pioneering AGB “star-in-a-box” models created by 
\cite{freyhoef08} and \cite {freyetal17}, 
building on the capability of the CO5BOLD code to cover the entire outer convective envelope and atmosphere, indeed show self-excited radial pulsations, with periods that are in good agreement with observations of Mira variables. 
\cite{liljetal18} 
indicated how the 3D model results can be used to improve the description of pulsation effects in 1D wind models.

\section{3D morphology of atmospheres and winds}\label{s_3Dwind}

A strong incentive to develop global 3D models of AGB stars and their dust-driven winds comes
from recent high-angular-resolution observations. Imaging of nearby AGB stars at visual and
infrared wavelengths has revealed complex, non-spherical distributions of gas and dust in the
close circumstellar environment,  
see, e.g., \cite{stewetal16}, \cite{wittetal17}. 
Temporal monitoring shows changes in both atmospheric morphology and grain sizes over the course of
weeks or months, 
e.g. \cite{khouetal16a, ohnaetal17}; see also the talk by Khouri, this conference. 
Such phenomena cannot be investigated with the spherically symmetric atmosphere and wind models mentioned above. In the 3D RHD “star-in-a-box” models by 
\cite{freyetal17}, 
on the other hand, an inhomogeneous distribution of atmospheric gas emerges naturally, as a consequence of large-scale convective flows below the photosphere and the resulting network of atmospheric shock waves. As shown in detail by 
\cite{hoeffrey19}, 
the dynamical patterns in the gas are imprinted on the dust in the close stellar environment, due to the density- and temperature-sensitivity of the grain growth process, explaining the origin of the observed clumpy dust clouds. However, these earlier 3D simulations did not include the effects of radiation pressure on dust, and could therefore not predict the structure of the wind formation zone. 

In Fig.\,\ref{f_3Dmod}, we show results of the first global 3D RHD “star-and-wind-in-a-box” models, computed with the CO5BOLD code (Freytag \& H{\"o}fner, in prep.). 
The models explore the interplay of interior dynamics (convection, pulsation), atmospheric shocks, dust formation, and wind acceleration in full 3D geometry. They include non-grey radiative transfer, as well as a time dependent description of silicate grain growth and evaporation, and they account for the effects of radiative pressure on dust in a simple way. These new models feature a much larger computational domain than previous “star-in-a-box” models, covering the inner wind region. This allows us to follow the emerging 3D structures to a distance where the outflow is established, and to compute mass loss rates.

\section{Summary and conclusions}

The winds of AGB stars are driven by radiation pressure on dust grains, which form in the highly dynamical atmospheres of these pulsating, strongly convective stars. Detailed 3D RHD models of the stars and their winds are essential for understanding the physical processes that cause the outflows, for explaining the observed properties of the stars, and for developing a predictive theory of mass loss on the AGB. 
High angular resolution imaging of scattered visual and near IR light shows clumpy dust clouds surrounding AGB stars. Such clumpy dust clouds emerge naturally in 3D “star-and-wind-in-a-box” models computed with the CO5BOLD code, as a consequence of giant convection cells and related atmospheric shock waves. 

Large dust grains with radii of about 0.1 -- 0.5 microns are found in scattered-light observations at distances of about 2 stellar radii around AGB stars, as required for driving winds by photon scattering on near-transparent silicate grains with low Fe/Mg ratios.  
New DARWIN models, describing the growth of olivine-type silicate grains with a variable Fe/Mg ratio, show that the enrichment with Fe is a secondary process, taking place in the stellar wind on the surface of Fe-free grains that have triggered the outflow. The self-regulating feedback between grain composition and radiative heating leads to low values of Fe/Mg, typically a few percent. These models show distinctive mid-IR silicate features even for low Fe/Mg, and realistic photometry from visual to mid-IR wavelengths. 

While 3D “star-and-wind-in-a-box” models are an essential tool for understanding the underlying physics of AGB star winds, they are computationally very demanding. For the foreseeable future, mass-loss descriptions for stellar evolution models will depend on extensive grids of 1D wind models, covering a wide range of stellar parameters. Recent results and ongoing theoretical work will help to constrain input parameters of such models, e.g. regarding stellar pulsation properties or dust nucleation. 

Finally, it should be mentioned that the wind mechanism of red supergiants is still unknown, but recent observations indicate that a significant part of the ejected mass is found in dense clouds of gas and dust. This could indicate similarities with AGB stars, regarding the mass-loss process (see talk by Chiavassa, this conference).

\begin{acknowledgements}
      The authors acknowledge funding from the European Research Council (ERC) 
      under the European Union’s Horizon 2020 research and innovation programme 
      (Grant agreement No. 883867, project EXWINGS) 
      and the Swedish Research Council (grant number 2019-04059). 
      The computations of 3D RHD models of AGB stars and their winds with the CO5BOLD code 
      were enabled by resources provided by the 
      Swedish National Infrastructure for Computing (SNIC).  
\end{acknowledgements}


\begin{thebibliography}{35}
\expandafter\ifx\csname natexlab\endcsname\relax\def\natexlab#1{#1}\fi

\bibitem[{Bladh} {et~al.} (2015)]{bladetal15}
{Bladh}, S., {H{\"o}fner}, S., {Aringer}, B., \& {Eriksson}, K. 2015, \aap,
  575, A105

\bibitem[{Bladh} {et~al.} (2013)]{bladetal13}
{Bladh}, S., {H{\"o}fner}, S., {Nowotny}, W., {Aringer}, B., \& {Eriksson}, K.
  2013, \aap, 553, A20

\bibitem[{Bladh} {et~al.} (2019)]{bladetal19}
{Bladh}, S., {Liljegren}, S., {H{\"o}fner}, S., {Aringer}, B., \& {Marigo}, P.
  2019, \aap, 626, A100

\bibitem[{Dorschner} {et~al.} (1995)]{dorsetal95}
{Dorschner}, J., {Begemann}, B., {Henning}, T., {Jaeger}, C., \& {Mutschke}, H.
  1995, \aap, 300, 503

\bibitem[{Eriksson} {et~al.} (2014)]{eriketal14}
{Eriksson}, K., {Nowotny}, W., {H{\"o}fner}, S., {Aringer}, B., \& {Wachter},
  A. 2014, \aap, 566, A95

\bibitem[{Freytag} \& {H{\"o}fner} (2008)]{freyhoef08}
{Freytag}, B. \& {H{\"o}fner}, S. 2008, \aap, 483, 571

\bibitem[{Freytag} {et~al.} (2017)]{freyetal17}
{Freytag}, B., {Liljegren}, S., \& {H{\"o}fner}, S. 2017, \aap, 600, A137

\bibitem[{Gail} {et~al.} (2016)]{gailetal16}
{Gail}, H.-P., {Scholz}, M., \& {Pucci}, A. 2016, \aap, 591, A17

\bibitem[{Gobrecht} {et~al.} (2016)]{gobretal16}
{Gobrecht}, D., {Cherchneff}, I., {Sarangi}, A., {Plane}, J.~M.~C., \&
  {Bromley}, S.~T. 2016, \aap, 585, A6

\bibitem[{Gobrecht} {et~al.} (2021)]{gobretal21}
{Gobrecht}, D., {Plane}, J. M.~C., {Bromley}, S.~T., {et~al.} 2021, arXiv
  e-prints, arXiv:2110.11139

\bibitem[{H{\"o}fner} (2008)]{hoef08}
{H{\"o}fner}, S. 2008, \aap, 491, L1

\bibitem[{H{\"o}fner} {et~al.} (2016)]{hoefetal16}
{H{\"o}fner}, S., {Bladh}, S., {Aringer}, B., \& {Ahuja}, R. 2016, \aap, 594,
  A108

\bibitem[{H{\"o}fner} {et~al.} (2021)]{hoefetal21arx}
{H{\"o}fner}, S., {Bladh}, S., {Aringer}, B., \& {Eriksson}, K. 2021, arXiv
  e-prints, arXiv:2110.15899

\bibitem[{H{\"o}fner} \& {Freytag} (2019)]{hoeffrey19}
{H{\"o}fner}, S. \& {Freytag}, B. 2019, \aap, 623, A158

\bibitem[{H{\"o}fner} \& {Olofsson} (2018)]{hoefolof18}
{H{\"o}fner}, S. \& {Olofsson}, H. 2018, \aapr, 26, 1

\bibitem[{J{\"a}ger} {et~al.} (2003)]{jaegetal03}
{J{\"a}ger}, C., {Dorschner}, J., {Mutschke}, H., {Posch}, T., \& {Henning}, T.
  2003, \aap, 408, 193

\bibitem[{Khouri} {et~al.} (2016)]{khouetal16a}
{Khouri}, T., {Maercker}, M., {Waters}, L.~B.~F.~M., {et~al.} 2016, \aap, 591,
  A70

\bibitem[{Liljegren} {et~al.} (2017)]{liljetal17}
{Liljegren}, S., {H{\"o}fner}, S., {Eriksson}, K., \& {Nowotny}, W. 2017, \aap,
  606, A6

\bibitem[{Liljegren} {et~al.} (2018)]{liljetal18}
{Liljegren}, S., {H{\"o}fner}, S., {Freytag}, B., \& {Bladh}, S. 2018, \aap,
  619, A47

\bibitem[{Norris} {et~al.} (2012)]{norretal12}
{Norris}, B.~R.~M., {Tuthill}, P.~G., {Ireland}, M.~J., {et~al.} 2012, \nat,
  484, 220

\bibitem[{Nowotny} {et~al.} (2013)]{nowoetal13}
{Nowotny}, W., {Aringer}, B., {H{\"o}fner}, S., \& {Eriksson}, K. 2013, \aap,
  552, A20

\bibitem[{Nowotny} {et~al.} (2011)]{nowoetal11}
{Nowotny}, W., {Aringer}, B., {H{\"o}fner}, S., \& {Lederer}, M.~T. 2011, \aap,
  529, A129

\bibitem[{Ohnaka} {et~al.} (2016)]{ohnaetal16}
{Ohnaka}, K., {Weigelt}, G., \& {Hofmann}, K.-H. 2016, \aap, 589, A91

\bibitem[{Ohnaka} {et~al.} (2017)]{ohnaetal17}
{Ohnaka}, K., {Weigelt}, G., \& {Hofmann}, K.-H. 2017, \aap, 597, A20

\bibitem[{Paladini} {et~al.} (2009)]{palaetal09}
{Paladini}, C., {Aringer}, B., {Hron}, J., {et~al.} 2009, \aap, 501, 1073

\bibitem[{Paladini} {et~al.} (2018)]{palaetal18}
{Paladini}, C., {Baron}, F., {Jorissen}, A., {et~al.} 2018, \nat, 553, 310

\bibitem[{Sacuto} {et~al.} (2011)]{sacuetal11}
{Sacuto}, S., {Aringer}, B., {Hron}, J., {et~al.} 2011, \aap, 525, A42

\bibitem[{Schwarzschild} (1975)]{schw75}
{Schwarzschild}, M. 1975, \apj, 195, 137

\bibitem[{Stewart} {et~al.} (2016)]{stewetal16}
{Stewart}, P.~N., {Tuthill}, P.~G., {Monnier}, J.~D., {et~al.} 2016, \mnras,
  455, 3102

\bibitem[{Trabucchi} {et~al.} (2017)]{trabetal17}
{Trabucchi}, M., {Wood}, P.~R., {Montalb{\'a}n}, J., {et~al.} 2017, \apj, 847,
  139

\bibitem[{Trabucchi} {et~al.} (2021)]{trabetal21}
{Trabucchi}, M., {Wood}, P.~R., {Mowlavi}, N., {et~al.} 2021, \mnras, 500, 1575

\bibitem[{Wittkowski} {et~al.} (2017)]{wittetal17}
{Wittkowski}, M., {Hofmann}, K.~H., {H{\"o}fner}, S., {et~al.} 2017, \aap, 601,
  A3

\bibitem[{Woitke} (2006)]{woit06}
{Woitke}, P. 2006, \aap, 460, L9

\bibitem[{Wood} (2015)]{wood15}
{Wood}, P.~R. 2015, \mnras, 448, 3829

\bibitem[{Zeidler} {et~al.} (2011)]{zeidetal11}
{Zeidler}, S., {Posch}, T., {Mutschke}, H., {Richter}, H., \& {Wehrhan}, O.
  2011, \aap, 526, A68

\end{thebibliography}
\end{document}